\documentclass[fleqn,usenatbib]{mn2e}
\usepackage{graphicx}
\usepackage{amsmath}
\usepackage{natbib}

\title{Realistic error estimates on kinematic parameters}
\author[V. De Bruyne et al.]
       {V. De Bruyne, P.Vauterin, S. De Rijcke, H. Dejonghe\\
        Astronomical Observatory, Ghent University, Krijgslaan 281, S9, 9000 Ghent, Belgium}
\date{Accepted 
      Received ;
      in original form }

\pagerange{\pageref{firstpage}--\pageref{lastpage}}
\pubyear{}

\begin{document}

\maketitle

\label{firstpage}

\begin{abstract}

Current error estimates on kinematic parameters are based on
the assumption that the data points in the spectra follow a Poisson
distribution.  For realistic data that have undergone several steps in
a reduction process, this is generally not the case. Neither is the
noise distribution independent in adjacent pixels. Hence, the error
estimates on the derived kinematic parameters will (in most cases) be
smaller than the real errors. In this paper we propose a method that
makes a diagnosis of the characteristics of the observed noise 
The method also offers the possibility to calculate
more realistic error estimates on kinematic parameters.
The method was tested on spectroscopic observations of NGC~3258. In
this particular case, the realistic errors are almost a factor of 2
larger than the errors based on least squares statistics. 
\end{abstract}


\section{Introduction}
The bulk of kinematic information available on elliptical galaxies is
nowadays retrieved through observation and analysis of the
line-of-sight velocity distributions (LOSVD's) of the stars in these
galaxies.

The basic idea behind the study of LOSVD's is that a galaxy spectrum
can be generated by the spectrum of a star of similar type that has
been redshifted by the rotation of the galaxy and smeared out by the
velocity dispersion of the stars (\citealt{min}).

The study of LOSVD's as we know it now started about 30 years ago,
when near the end of the seventies, \citet{sar} and \citet{ton}
published surveys of galaxy redshifts and velocity distributions. The
methods they presented in their papers were widely used afterward:

\begin{itemize}
\item Fourier quotient method \citep{sar}: The LOSVD is obtained by
dividing the Fourier transform of the galaxy spectrum by the Fourier
transform of the template spectrum, and transform the result back into
pixel space.  Although it seems to be natural to use Fourier space,
this has also some disadvantages. The error calculation becomes
complicated due to the correlations in the quotient and the method is
very sensitive to template mismatch.

\item Cross-correlation method \citep{ton}: The peak in the
galaxy-template correlation function is fit by a Gaussian
function. The method also provides a way to obtain error estimates on
the kinematic parameters.
\end{itemize}

These techniques assume a Gaussian form for the broadening functions
for mainly practical reasons: it provides a way of filtering the noise
in the spectra and it allows to calculate convolutions analytically.
The mean and variance determining the Gaussian function were
considered as parameters describing the mean streaming and stellar
velocity dispersion.

About ten to fifteen years later, the quality of the data had improved
a lot and new techniques were developed trying to get more information
out of the observed spectra than just two parameters.  The main
improvement was the elimination of the assumption of a Gaussian form
for the LOSVD. There are mainly two criteria that can be used to
classify these techniques: (a) whether a
non-parametric LOSVD or a parameterized LOSVD is derived. In case of
high S/N data, a non-parametric LOSVD yields all information you can
get out of a spectrum. But because of the presence of noise there is
always a need for some sort of smoothing.  There are several ways to
do this, either by adapting some filter or by using a smoothing
parameter. For these non-parametric fitting methods there is not
always a simple way to come to an error estimate. (b) whether the fit
is performed in Fourier space or in pixel space.  A treatment of the
problem in pixel space has the advantage that parts of the spectrum
can be easily eliminated and allows an easier error estimate.

A lot of methods have been proposed:
\begin{itemize}
\item Fourier fitting method \citep{f1}: (parametric in Fourier space)
A convolution of the LOSVD with the template is fitted to the galaxy
spectrum in Fourier space. Error estimates are derived from a least
squares fit.

\item Fourier correlation quotient method \citep{b1}: (non-parametric
in Fourier space) The method is based on the deconvolution of the peak
of the template-galaxy correlation function with the peak of the
autocorrelation of the template star.  This approach is shown to be
less sensitive to template mismatch than the standard Fourier quotient
method. The deconvolution itself is done in Fourier space. Noise
filtering is done by applying a Wiener filter. Error estimates can
be retrieved from Monte Carlo simulations.

\item Direct fitting method \citep{r1}:
(non-parametric in pixel space)
The galaxy spectrum is assumed to be a superposition of various
components: (1) continuum terms, (2) a set of broadened stellar
spectra where the broadening function and the stellar spectra can
vary, (3) a noise component. The best combination of components and
the errors on the results are obtained from a least squares
minimization.

\item unresolved Gaussian deconvolution \citep{k1}:
(non-parametric in pixel space) The LOSVD is modeled as a sum of a set
of Gaussian distributions that are uniformly spaced in
velocity. 
The best fit is determined by
means of a least square minimization, which also serves as basis for
  error determination.

\item Bayesian method \citep{s2}: (non-parametric or parametric in
pixel space) The concept is to take a trial LOSVD and convolve it with
a template spectrum into a model spectrum. If this model is the real
one, the difference between the model spectrum and the observed
spectrum must be the noise. With probability theory it can be
calculated what the probability is that the trial LOSVD is the true
one, given the data. Confidence intervals for broadening functions are
obtained from Monte Carlo simulations.

\item Cross-correlation method update \citep{st}: (parametric in pixel
space) This is a generalization of the galaxy-template
cross-correlation method to arbitrary parameterized line profiles,
where the correlation function is fitted by the model line profile
using a least squares technique. For a well resolved galaxy spectrum,
the result is fairly insensitive to template mismatch. Errors
follow out of standard statistics involved in the fitting technique.

\item penalized likelihood \citep{m1}: (non-parametric in pixel space)
The method is based on a penalty function that is large for any noisy
function and zero for a Gaussian function. This method allows to fit
high quality data very accurately, whereas also very noisy data can be
fitted, in the limit by a simple Gaussian. Monte Carlo simulations are
performed to come to an error estimate.
\end{itemize}

When parameters based on the integrated moments of the LOSVD are
sought for, one can also obtain these parameters in a one step
process. In this case, a convolution of a parameterized expression for
the LOSVD with a stellar template spectrum is matched to the observed
galaxy spectrum.

\begin{itemize}
\item Gauss-Hermite fitting method \citep{vm}:
(parametric in Fourier space) A
Gauss-Hermite parameterization is used for the LOSVD, a least squares
minimization is performed in Fourier space.

\item Gauss-Hermite direct fitting method \citep{vm2}: 
(parametric in pixel space) The LOSVD
is written as a truncated Gauss-Hermite series and the least squares
fit is performed in pixel space.
\end{itemize}

It is generally believed that, if the data are good enough to apply
non-parametric fitting, such an approach implies less bias than a
parametric fit. However, having a non-parametric LOSVD, it is still
very useful to describe the obtained profile by just a few
characteristic numbers. Higher order information can be obtained
either by calculating the moments of the velocity distribution or by
fitting a parameterized function to it.  Since there is a real danger
of confusion when dealing with the moments of the LOSVD, we define
them here explicitly:
\begin{equation}
 \langle v_p\rangle = \int dv_p v_p {\cal L}(v_p),
\end{equation}
and

\begin{equation}
\mu_k = \int dv_p (v_p-\langle v_p\rangle)^k {\cal L}(v_p),\quad k\ne1,
\end{equation}
with ${\cal L}(v_p)$ the LOSVD.  The kinematic moments are the mean
projected rotation $\langle v_p\rangle$, the projected
velocity dispersion $\sigma_p^2 = \mu_2$, the skewness $\zeta_3
= \mu_3/\sigma^3_p$ and the kurtosis $\zeta_4 = \mu_4/\sigma^4_p$.

The calculation of higher order moments out of LOSVD's may be
troublesome because of the strong dependence of the moments on the
wings of the profiles.  These wings unfortunately are the hardest to
retrieve from observations. Therefore, in most practical cases, a
parameterized function is fitted to the LOSVD. A truncated
Gauss-Hermite series is often used (\citealt{g1},\citealt{vm}):
\begin{equation}
{\cal L}(v_p) = \gamma\exp\left[-{1\over 2} {(v_p - \langle v\rangle)^2 \over \sigma^2}\right] \times \left[1+\sum_{j=3}^n h_j H_j\left({v_p - \langle v\rangle \over \sigma}\right)\right].
\end{equation}

In this expression, $\langle v\rangle$ and $\sigma$ are free
parameters, they are also called 'mean velocity' and 'velocity
dispersion'. The functions $H_j(x)$ are the $j$-th order Gauss-Hermite
polynomials, and their coefficients $h_j$ are regarded as the higher
order parameters.  The coefficient $h_3$ is the third order shape
parameter and is an indication for anti-symmetric deviations from a
pure Gaussian function in the form of leading or trailing tails. The
coefficient $h_4$ is the fourth order shape parameter and is an
indication for symmetric deviations from a pure Gaussian LOSVD in the
form of broader or smaller wings.

Kinematic information up (to parameters corresponding) to the fourth
order moment of the LOSVD is becoming available for a large sample of
ellipticals, see e.g \citet{bsg}, \citet{kr1}, \citet{h1}.  These data
sets are a prerequisite for the construction of dynamical models of
elliptical galaxies. In some cases it can be noticed that the error
estimates on the parameters are smaller than the scatter among
neighbouring data points in the kinematic profiles. This is especially
striking for the so-called higher order moments (third and fourth
order moment). Examples of this behaviour can be seen in some of the
data presented by \citet{bsg}, \citet{kr1}, \citet{h1}.  A
possible interpretation of this fact is that the real errors on the
kinematic parameters are larger than the error estimates given by the
parameter fitting algorithms.

The derivation of these error estimates relies on assumptions that are
generally not completely met. Though people are aware of this, it is
generally unclear how large the discrepancy between reality and
assumptions is.

One can come up with at least two good reasons why it is important to
have realistic error estimates for dynamical modelling.  First, for
the construction of dynamical models, it is convenient to work with a
goodness of fit indicator, which is meant to be an objective indicator
of how good the model fits the data.  Such goodness of fit indicators
need error estimates on the data.  However, if these are not realistic
errors, it is not straightforward to interpret the values of this
goodness of fit indicator, let alone derive meaningful confidence
intervals for the models. In some cases, awareness of this problem may
prompt one to adopt modified error estimates for use in the goodness
of fit indicator (e.g. \citet{ko1}, \citet{cr1}, \citet{g2},
\citet{s1}, \citet{kr1}). However, playing with the error estimates on
individual data points is equivalent to using different relative
weights for the data points, and this is likely to complicate the
interpretation of a goodness of fit indicator rather than simplify it.

A second issue of concern is that these parameters are often used to
determine a realistic dynamical model and to constrain the potential
of a galaxy at the same time. In particular the velocity dispersion
and the fourth order parameter play an important role in inferring the
existence of central black holes or dark matter haloes around
elliptical galaxies. In practice this is done by assuming a potential
for the system and trying to find a dynamical model based upon the
selected potential that simultaneously fits the velocity dispersion
and fourth order parameter within the given error estimates. If no
such model can be found, it is concluded that the assumed potential is
not the right one, and the amount of dark matter is changed. If the
  error estimates are smaller than the real errors and
the conclusions on the potential critically depend on them (e.g.
\citet{r2}, \citet{kr1}, \citet{cr2}), some potential-model combinations
may be rejected wrongly.  As an inevitable consequence, it is possible
that in some cases the use of realistic error estimates in dynamical
modelling will weaken the dynamical evidence for dark matter in
elliptical galaxies.

The method to derive realistic error estimates on kinematic parameters
that is presented in this paper reaches out toward a need on the
observational side and the modelling side. The method makes an
implicit diagnosis of the noise distribution and uses the
characteristics of this distribution as the basis of a new error
estimate.

In section \ref{diag}, a diagnosis of the problem with error estimates
is given.  An outline of the new method is presented in section
\ref{outline}, this is elaborated and illustrated in section
\ref{gal}. A discussion and conclusions can be found in sections
\ref{disc} and \ref{conc} respectively.

\section{Diagnosis of the problem}\label{diag}

In the absence of external sources of errors, a raw
spectral image contains data points that follow a Poisson
distribution, the variance of which is denoted by "Poisson noise". 
Another characteristic of a raw spectral image is that the data points
in the image are independent.  Hence, the power spectrum of the raw
signal would match the definition of white noise: it has constant
power in the frequency domain.

Before kinematic parameters can be calculated, the spectra have to be
brought in a state that is referred to as 'cleaned' and
'calibrated'. Bringing the spectra in such a condition requires a
number of image processing steps, the so-called 'standard reduction
techniques', for which adequate tools are available in image analysis
packages such as MIDAS or IRAF. But besides bringing the observed
spectra in good condition, these reduction steps also have a
non-negligible influence on the noise distribution.

For example, the removal of bad pixels or cosmic ray events, where the
values of two or more adjacent pixels are replaced by an average over
a surrounding region, clearly introduce dependencies in the data.
Likewise, dependencies are introduced by any operation where some kind
of interpolation is involved, like e.g. calibration.
Moreover, a sky subtraction does certainly not guarantee a perfect
elimination of the sky in the data, regardless the level of
sophistication that is used to perform the operation.  The residual of
this correction also adds up to the noise distribution.  In some cases
the above effects can be the source of considerable uncertainties on
the derived kinematic parameters, while this is not reflected by error
estimates based on Poisson noise.

In many cases, a least squares technique is used to derive the
kinematic parameters and error estimates from the data.  The
statistical interpretation of this method relies on the assumption
that (1) the noise is independent and (2) Gaussian distributed on the
input data. For large numbers, a Gaussian distribution is a fairly
good approximation for a Poisson distribution and therefore the second
assumption is valid. It is clear from the above arguments that the
first condition generally is not met after data reduction. As a
consequence, the errors derived from standard statistics will in many
cases differ from the real errors on the kinematic parameters.

Sometimes, Monte Carlo simulations are used (\citet{bsg}, \citet{st})
to estimate the uncertainties on the derived parameters. For the
realization of synthetic galaxy spectra, a Gaussian noise distribution
with given S/N is used. So also this method may show a similar
tendency to underestimate the errors.

Accepting the fact that the cleaned and calibrated spectral images
come as they are, it is possible to obtain error estimates that are
more realistic estimates than those currently found in
literature. This is the purpose of the present paper.

\section[]{Outline of the method}\label{outline}
To get an idea of more realistic errors on the
parameters and to overcome the above problems we propose the following
scheme. The steps are described in more detail in the following
sections.
\begin{itemize}

\item Perform a fit of a model galaxy spectrum to the observed galaxy
spectrum and determine the residual of this fit.  The model galaxy
spectrum is a convolution of the selected template spectrum with a
LOSVD.  For this step it is important to use an expression for the
LOSVD that offers enough freedom. 
The purpose is to fit every part of the galaxy spectrum that can
reasonably be fit by a convolution of the template spectrum with a
general and smooth LOSVD.

\item The residual of this fit contains (1) the Gaussian noise on the
raw spectrum, (2) the features left or introduced by several reduction
steps. The errors estimated with the proposed method take into account
artifacts from flatfielding, cosmic remnants, effects of rebinning and
remnants of sky removal, for as far as the latter has been removed
reasonably well. On the other hand, the method does not account for
errors in continuum subtraction nor template mismatch. 
Hence, the residual of the fit can be considered as a realization of
the "observed noise" involved in the problem: the term "observed noise"
applies to 'cleaned and calibrated' spectra and is used to make a
distinction with Poisson noise. Where pure white noise is
featureless and transforms into a flat power spectrum, the power
spectrum of the residual will be different.
A smooth representation of the power spectrum of the residual will be
taken as sufficient information on the main characteristics of the
observed noise.

\item The next step is to generate a number of synthetic galaxy
spectra, that have a noise distribution with the same power spectrum as
the observed noise, and to use these synthetic spectra to determine
the spread on the kinematic parameters.  These equivalent noise
distributions that carry the characteristics of the original observed
noise can be obtained by multiplying the Fourier transform of a white
noise profile with a smooth representation of the power spectrum of
the residual, and transforming this back.
\end{itemize}

The idea of using Monte Carlo simulations to determine error estimates
of course is not new, but the main point is that in the present method
one tries to model the real characteristics of the noise.

A demonstration of the method on realistic galaxy spectra is given in
section \ref{gal}.

\section[]{NGC~3258}\label{gal}

\subsection{Observations}
NGC~3258 is an E1 galaxy and was observed with the ESO-NTT telescope
in the nights of 27-28/2/2001 (64.N-192). Spectra of the major axis
were taken using the red arm of EMMI, covering the Ca II
triplet. Grating \#7 was used, having a dispersion of 0.66 \AA/pix.
The detector was a Tektronix CCD with 2048$\times$2047 pixels, 24$\mu
m \times$24$\mu m$ in size and with a pixel scale of $0.27''$/pixel.
A slit width of $1.5''$ thus yields a spectral resolution of 3.67
{\AA} FWHM, resulting in an instrumental dispersion of about 54 km/s
in the region of the Ca II triplet.  For NGC~3258, several exposures
of 3600 sec were taken (in total 11 hours).  A number of standard
stars (G dwarfs and K and M giants) were also observed.

\subsubsection{Standard data reduction steps}

Standard reduction steps were applied to these spectra.  Out of the
bias and dark exposures a correction term was derived.  Several
domeflat images were used to come to an appropriate flatfield image.
Cosmics were removed using a top hat filter. After filtering, the
images were inspected carefully and remaining cosmics were also
removed by hand.

For the wavelength calibration, lamp spectra were taken just before or
after each of the spectroscopic observations of the galaxy or the
stellar template stars. For each row in the lamp spectra images, a
polynomial was constructed to transform the pixel scale into
wavelength scale. The same coordinate transformation was then applied
to the galaxy or star spectra. The calibrated images have a step of
$0.3$\AA.

For the airmass correction, the mean value of the airmass at the
beginning and end of the exposure was used.  The contribution of the
sky to the spectra was estimated from an upper and lower region of the
image, where there was no contribution of the light of the galaxy or
template star. Several galaxy spectra were reduced separately, aligned
and combined into one galaxy spectra image.

Spatial rebinning of the galaxy spectra resulted in data with a S/N
$\ge$ 50/bin (about 70 in the centre).
Only a small part of the spectrum, from 8500 {\AA} to 8750
{\AA} was used to determine the kinematic parameters.

\subsection{The observed noise on the data}

The idea is that the residual of the observed galaxy spectrum with a
sufficiently flexible model spectrum can be attributed to some sort of
noise involved in the problem.  For a given stellar template spectrum
it is important to consider a representation for the LOSVD that offers
enough degrees of freedom to be able to find the best fitting model
spectrum.

For this reason, nonparametric LOSVD's in combination with a flexible
method to generate them are chosen to perform this fit. We have also
chosen not to restrict ourselves to positive LOSVDs, because this
makes the fit more general and at this point there are no physical
restrictions on the LOSVDs. The method adopted here performs a least
squares fit, using a set of cubic spline basis functions to represent
the LOSVD.  An outline of the method can be found in Appendix
A. However, for this step, any non-parametric fitting method with
enough degrees of freedom will do.  The model spectrum was a linear
combination of convolutions with a number of observed template stars
in which the coefficients were chosen on the basis of a goodness of
fit estimator.

The non-parametric fit is only used to get an idea of the fraction of
the signal in the galaxy spectrum that cannot be accounted for by a
model spectrum with the chosen template mix, and hence that will be
part of the residual. Furthermore, also traces of sky lines, or
incompletely removed cosmics will contribute to the residual of this
fit.

\begin{figure}
\includegraphics[scale=.5]{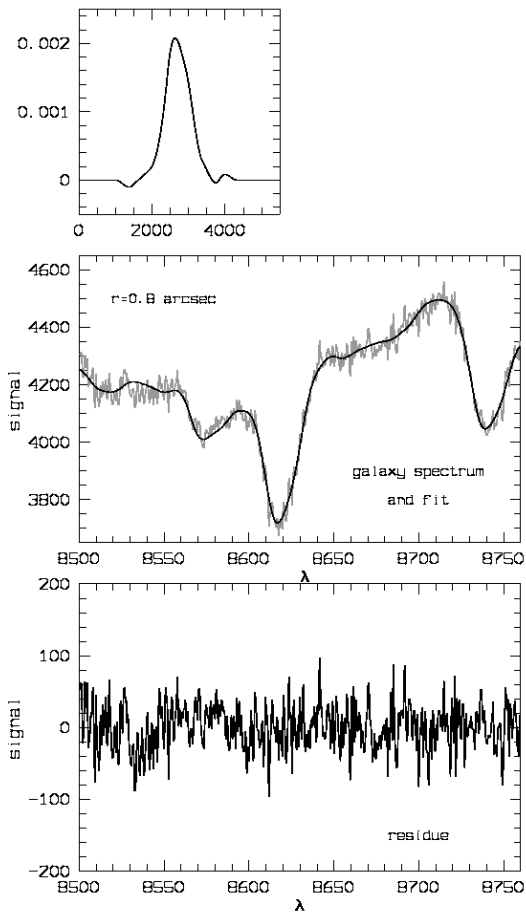}

\caption{At position $r=0.8''$: the galaxy spectrum and non-parametric
fit in the middle panel and the residual of the fit in the lower panel.
The upper panel shows the LOSVD resulting from the non-parametric fit.}

\label{pos2}
\end{figure}

\begin{figure}
\includegraphics[scale=.5]{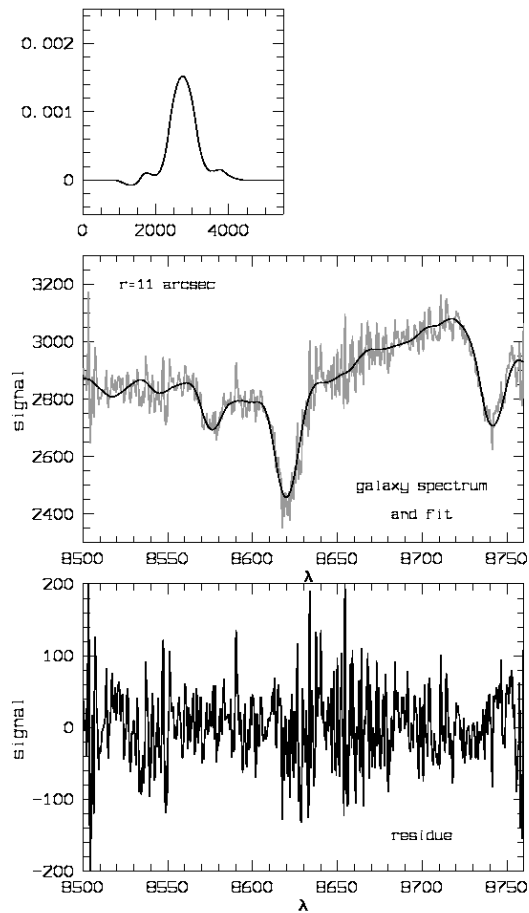}

\caption{At position $r=11''$: the galaxy spectrum and non-parametric
fit in the middle panel and the residual of the fit in the lower
panel.  The upper panel shows the LOSVD resulting from the
non-parametric fit.}

\label{pos9}
\end{figure}

As an illustration, the results of the method for two different
positions along the major axis of NGC~3258 are shown in figure
\ref{pos2} for $r=0.8''$ and figure \ref{pos9} for $r= 11''$.  The
upper panel displays the LOSVD, obtained from a non-parametric fit
with 10 degrees of freedom. The middle panel shows the observed galaxy
spectrum and the fit. The lower panel shows the residual of the fit.
From these lower panels is it clear that the residual at $11''$
(values mostly between -100 and 100) is much larger than the residual
at $0.8''$ (values mostly between -60 and 60). This immediately shows
that the Poisson noise is not the only source of errors. The galaxy
spectrum at $0.8''$ has much more counts in absolute value than the
galaxy spectrum at $11''$, hence pure Poisson noise is higher at
$0.8''$ than at $11''$. The surplus non-Poissonian contribution to the
residual at $11''$ is likely to come mostly from an imperfect sky
subtraction.

\subsection{The power spectrum of the residual} 

\begin{figure}
\includegraphics[scale=.5]{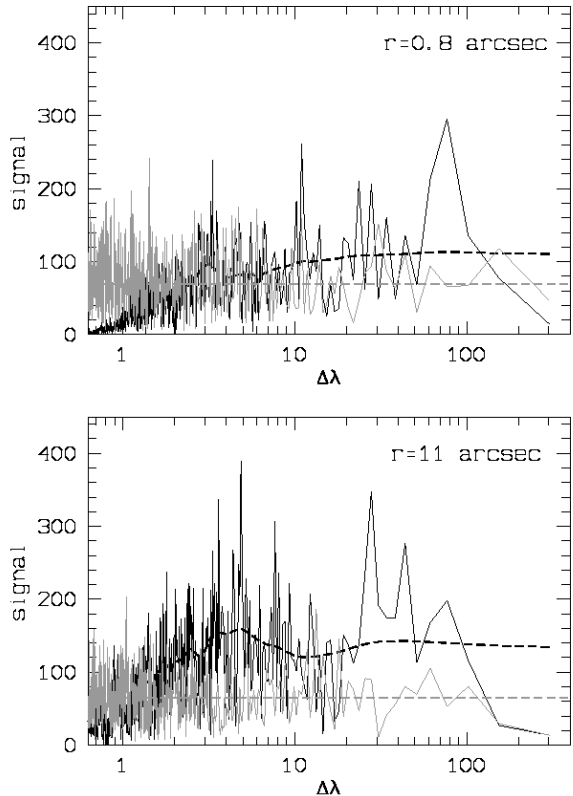}

\caption{The power spectrum of the noise involved in the fit (black
line) and white noise (gray line). The smoothed observed noise profile is
shown in black dashed line, the constant level of the white noise is
shown in gray dashed line. Upper panel: at position $r=0.8''$.  Lower
panel: at position $r=11''$.}

\label{powers}
\end{figure}
To see what the magnitudes of the various frequency components in
these residuals are, their power spectra have been calculated.  They
are presented in figure \ref{powers}, the residual at $0.8''$ is in the
upper panel, the residual at $11''$ in the lower panel.  The scale on
the horizontal axis is the period of the power spectrum, instead
of the more widely used frequency. This period (labelled $\Delta
\lambda$) is directly related to the wavelength scales of the features
in the original spectrum. Because of the wide range in values for
$\Delta \lambda$, a logarithmic axis is plotted.  On top of the
residual power spectra in figure \ref{powers}, a sample power spectrum
for the expected white noise at these radii is presented. There is a
clear difference between both profiles.

Having these power spectra representations, one would like to know
which part is playing an important role for the determination of real
errors on the kinematics. In other words, which frequencies can have
an important impact on the derived kinematic parameters?

\begin{figure}
\includegraphics[scale=.5]{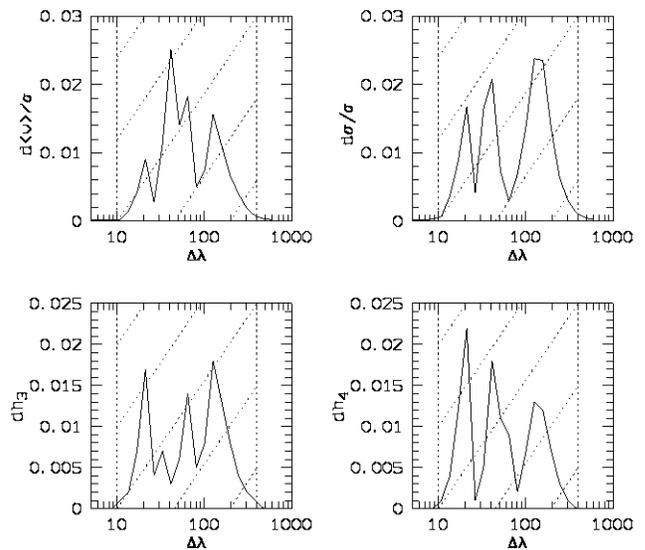}

\caption{The shaded regions show the regions of influence of injected
wave functions with different wavelengths on the measurement of
kinematic parameters.}

\label{freq}
\end{figure}

This can be investigated by means of some simulations.  Galaxy spectra
with various noise characteristics can be created starting from a
galaxy spectrum obtained after convolution of a template spectrum with
a known LOSVD and adding wave functions (sine and cosine functions)
with different frequencies.  For these artificial spectra, a best
fitting model spectrum was determined using a least-squares
minimization. The model spectrum was composed with the template
spectrum and a LOSVD expressed as a truncated Gauss-Hermite series.
This analysis of the artificial spectra gives new values for the
kinematic parameters, that will differ from the original kinematic
parameters.  The results of such simulations as a function of the
scale size of the waves can be seen in figure \ref{freq} (upper panels
for d$\langle v\rangle$/$\sigma$ and d$\sigma$/$\sigma$, lower panels
for d$h_3$ and d$h_4$).
 
It is clear that the measurement of these kinematic parameters is
sensitive to superpositions of wave functions with wavelengths
($\Delta \lambda$) between 10 {\AA} and 400 {\AA} (hashed regions in
figure \ref{freq}). This $\Delta\lambda$ range is independent of the
wavelength range used in the analysis, but scales with $\sigma$ (which
is 325 km/s in this case). 

This learns that the region of interest in the power spectrum of the
residuals lies between $\Delta\lambda=10${\AA} and
$\Delta\lambda=400${\AA}. It is also there (see figure \ref{powers})
that the residual power spectra overshoot the white noise.

For short wavelengths, the residual power spectrum seems to have lower
values than the Poisson noise power spectrum. At first sight this
seems to be puzzling. However, this is a result of the
interpolations that took place in the course of the data reduction
process. Interpolation always implies some sort of smoothing, which
means that features with a small wavelength scale disappear from the
image.

Moreover, oversampling lies at the origin of the transfer of an amount
of noise coming from small wavelength scale to larger wavelength scale.
The stronger the oversampling, which has the artificially enlarging of
feature wavelength scales as a result, the stronger this migration
effect.  In the case of NGC~3258, the original spectrum had a step of
$0.66$\AA/pixel, while the spectra after calibration have a step of
$0.3$\AA/pixel.

\subsection{Realistic error estimates}
The power spectra of the residuals clearly show that the assumption of
independent (white) noise is not a valid one. This implies that the
classical statistical tools for calculating error estimates out of
least squares fitting techniques cannot be used.

The only option that is left, is to determine the uncertainties on the
parameters through Monte Carlo simulations. In this case it is
important to work with simulated galaxy spectra that have the same
noise characteristics as the original galaxy spectrum. Our method
proposes to achieve this by using simulated galaxy spectra that give
residuals that follow the same power spectrum as the initially
determined residual.

To realize this, a smooth representation of the residual power spectrum
is created. This is illustrated in figure \ref{powers}, by means of
the dashed lines. In this case, each pixel was replaced by a weighted
mean over 20 pixels at each side taking a linearly declining weight
function into account. This smooth function has basically the same
behaviour as the power spectrum, and hence carries the information
that distinguishes the observed noise distribution from white noise.  New
realizations of this noise distribution can be calculated by
multiplying a white noise spectrum with this fitted function.  In this
way, the simulated galaxy spectra for the Monte Carlo simulation are
constructed.

To these spectra, a model spectrum composed with the template spectrum
and a parameterized expression for the LOSVD is fitted. For the
parameterization, a Gauss-Hermite decomposition was used, following
\citet{vm}. The best fitting values for $\langle
v\rangle$, $\sigma$, $h_3$ and $h_4$ were obtained using the
Levenberg-Marquardt method as described in \citet{p1}.

\begin{figure}
\includegraphics[scale=.55]{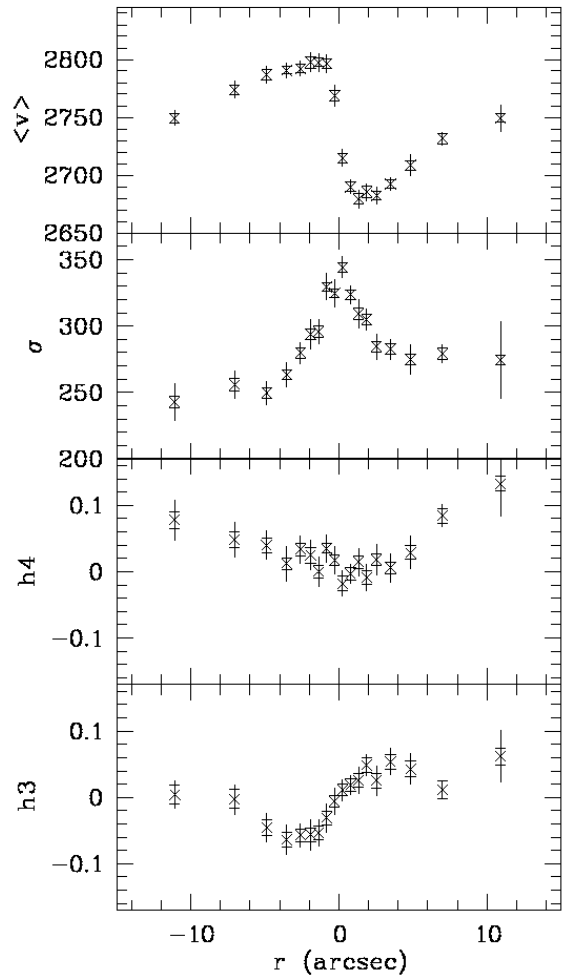}
  \caption{Kinematic parameters for NGC~3258. The realistic error estimates are shown by means of vertical lines, the horizontal lines on these error estimates indicate an estimate for the errors based on least squares statistics.}
 \label{kinemat}
\end{figure}

The error estimates are the root mean squares of the kinematic
parameters that were derived from 100 realizations of the galaxy
spectra. The results are shown in figure \ref{kinemat}, from upper
to lower panel: the mean rotation velocity, the velocity dispersion,
$h_4$ and $h_3$. The vertical error estimates present realistic
errors. The horizontal lines correspond to error estimates based on
least squares statistics.

The mean velocity reaches a maximum value of about 56 km/s and the
realistic error estimates lie between 5.7 km/s and 11.3 km/s (with a
mean error of 7.5 km/s and a mean error from least squares statistics
of 4 km/s). The central velocity dispersion is 344 km/s, with an error
of 8 km/s.  The mean error on $\sigma$ is 10.4 km/s, the mean error on
$\sigma$ based on least squares statistics is 3.8 km/s.  For large
radii, the profile is antisymmetric. 

The profile for $h_3$ goes through the origin, apparently there is no
serious problem with template mismatch. 
The mean error on $h_3$ is 0.02, the lowest value is about 0.013, the
highest value is 0.039 (measured in an outer point). From least squares
statistics, the mean error on $h_3$ is 0.01.

The $h_4$ profile is approximately symmetric, with low values in the
centre and higher values for the outer data points. The central data
points show quite some scatter, but this is not inconsistent with the
errors (at least 0.014) on the data in that region. The errors increase
outwards, the outer data points have an error of 0.049 (positive side)
and 0.03 (negative side). The mean error on $h_4$ is 0.023, the mean
error on $h_4$ following least squares statistics is 0.01.

For the $h_3$ and $h_4$ profiles, we compared the data points at
either side of the centre. The amount of scatter in the profiles can
be expressed as the mean distance between corresponding data points at
positive and negative radius, $\sqrt{(\sum_{i=1,n} (|y_i|-|y'_i|)^2)/n}$,
with $n$ the number of data points at positive or negative radius, and
$y_i$ and $y'_i$ the values for the parameters in corresponding points
at positive and negative side of the centre.  Fore $h_3$ the value of
this scatter indicator is 0.019 and for $h_4$ this is 0.031.  This
means that for both parameters, the mean errors derived with the
proposed method are close to this scatter indicator, while the mean
errors following from least squares statistics are clearly smaller.

\section[]{Discussion}\label{disc}

To verify and validate the algorithm and its implementation, we now
explore a few situations in which (following the discussion in section
\ref{diag}) conventional and more realistic error bars based on the
method described in this paper are expected to give different results
by means of simulations. Using simulations has the advantage that also
the true errors can be calculated and hence the error estimates can be
compared.

\subsection{Impulse noise}
A fraction of the noise with non-Poissonian character is manifested as
spikes that overshoot the Poisson noise and that are relatively small
on wavelength scale. This situation can be mimicked in a simulation by
adding impulse noise to a spectrum with otherwise pure Poisson noise.

Synthetic model spectra were obtained by convolving a LOSVD
characterized by $\langle v\rangle = h_3 = h_4=0$ and $\sigma=250$
km/s with a K3 giant and adding Poisson noise.
Different types of model spectra were then created by
contaminating 10 $\%$ of the pixels in the spectrum with impulse noise
with an amplitude of 3, 6, 9, 12 or 15 times the amplitude of the
Poisson noise.

\begin{figure}
\includegraphics[scale=.5]{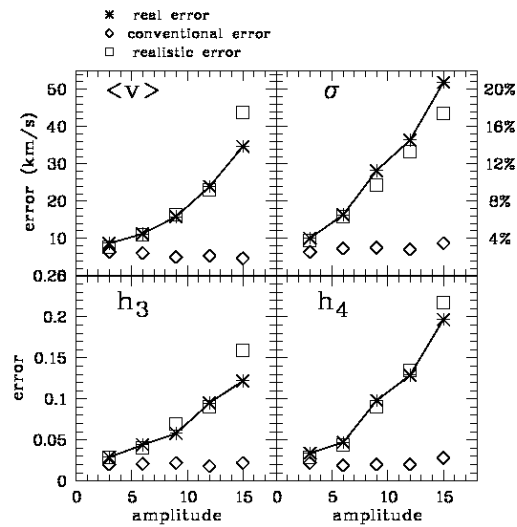}

\caption{Results of a simulation where impulse noise is added to a
spectrum with pure Poisson noise. The amplitude of the impulse noise,
relative to the amplitude of the Poisson noise is indicated on the
horizontal axis. On the vertical axis are the measured errors. The
asterisks indicate the true errors on the kinematic parameters, the
squares indicate the realistic errors and the diamonds indicate the
conventional errors.}

\label{spike}
\end{figure}

The RMS deviation on the kinematic parameters calculated out of 100
equivalent realization of one type of synthetic spectrum are
considered to be the true errors on the kinematic parameters in that
specific situation.  For each type of spectrum, also the analyses
following the scheme outlined in section \ref{outline} was applied
(these results will be referred to as realistic errors). Likewise,
conventional errors were calculated with a standard least squares technique.
These results are presented in figure \ref{spike}. The true errors are
indicated in asterisks, the realistic errors in squares and the conventional
errors in diamonds.

It is clear that the conventional errors returned by a standard technique
are independent of the degree of contamination and can underestimate
the true errors considerably. For an impulse noise with an amplitude
of 3 times that of the Poisson noise the conventional error underestimates
the true error by $\sim 27 \%$ for $\langle v \rangle$, $\sim 37\%$
for $\sigma$, $\sim 31\%$ for $h_3$ and $\sim 35\%$ for $h_4$. The
realistic errors on the other hand follow the true errors much
better. For the same situation the difference between realistic errors
and true errors is $\sim 10 \%$ for $\langle v \rangle$, $\sim 5\%$
for $\sigma$, $\sim 0\%$ for $h_3$ and $\sim 14\%$ for $h_4$. For
serious contamination, with an impulse noise with an amplitude of 15
times that of the Poisson noise, the realistic error bars seem to
slightly overestimate the real errors, but this is hopefully no longer
a realistic situation.

\subsection{Sampling}

Changing the original sampling step of a spectrum has an impact on the
error propagation that is completely neglected when Poisson noise is
used as basic assumption for the error calculation.

For the simulations, a number of synthetic spectra were obtained with
the same LOSVD and template star as above, but the number of pixels
was different for each spectrum. In the end, all synthetic spectra
were resampled in order to end up with spectra with $N$ pixels. This
means that for the final spectra, the wavelength scale was $n$
times the original wavelength scale, with $n$ equal to 1.5, 2, 3 or 4,
in order to mimic oversampling, and also $n$ equal to 0.5 and 0.25,
thus mimicking undersampling.

\begin{figure}
\includegraphics[scale=.5]{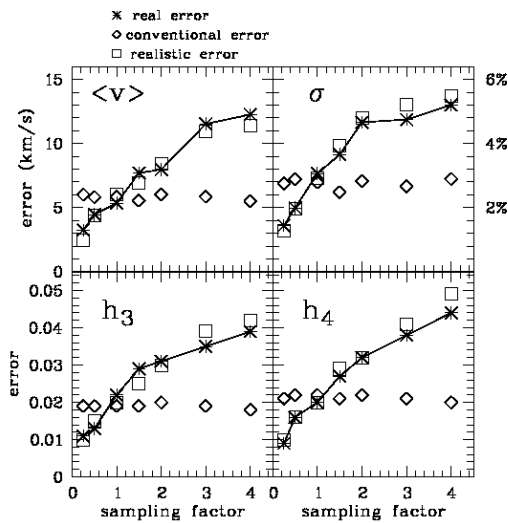}
\caption{The results of simulations with rebinned spectra, where the
wavelength scale after rebinning is $n$ times the wavelength scale of
the original spectrum. The sampling factor $n$ is on the horizontal
axis, the vertical axes represents the errors. The true errors are
indicated in asterisks, the realistic errors in squares and the conventional
error in diamonds.}
\label{sample}
\end{figure}

These sampling factors are chosen only for the sake of illustrating
the method.  These rebinned spectra were analysed using the technique
described in section \ref{outline}, yielding realistic errors, and
using a standard technique for the conventional errors. Again, the real
errors were calculated out of Monte Carlo simulations with 100
spectra. The resulting error estimates are presented in figure
\ref{sample}, in squares for the realistic errors and diamonds for the
conventional errors. The true errors are indicated in asterisks.  Again, the
conventional errors are independent of any change in amount of over- or
undersampling, while the realistic errors mainly follow the true
errors.  The change in errors scales roughly with $\sqrt n$, with $n$
the sampling factor.

This result shows that care should be taken if kinematic parameters
and conventional errors are estimated from spectra that are logarithmically
rebinned. In that case the wavelength scale is clearly changed, but in
such a way that some regions are more compressed and other regions are
stretched.

\subsection{Template mismatch}
\begin{figure}
\includegraphics[scale=.5]{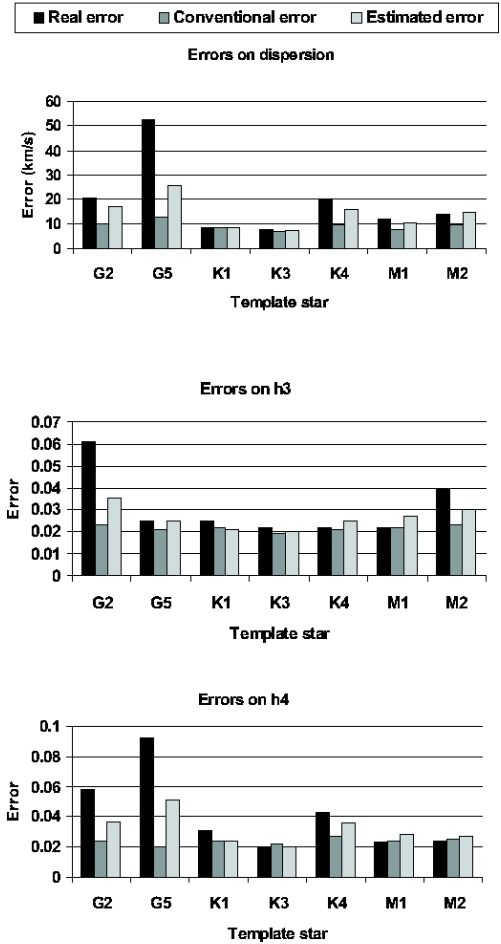}
\caption{The results of simulations with other templates. The horizontal axis shows the sequence of template stars,
the bars indicate the errors: black for the true error, dark gray for
the conventional error and light gray for the "realistic" error using this
technique. The results for $\sigma$ are in the upper panel, the
results for $h_3$ in the middle panel and the results for $h_4$ in
the lower panel.}
\label{temp}
\end{figure}

We do not claim that the proposed method is able to cope with the
errors coming from template mismatch. But also in this case, the
errors estimated with the method presented in this paper are closer to
the true errors than the conventional errors obtained by standard
techniques.  The synthetic spectrum, created with a K3 giant, was also
analysed with the following templates: G2 dwarf, G5 dwarf, K1 giant,
K4 giant, M1 giant, M2 giant.  The results are presented in figure
\ref{temp}. The horizontal axis shows the sequence of template stars,
the bars indicate the errors: black for the true error, dark gray for
the conventional error and light gray for the realistic error
(indicating that the error is obtained using the technique presented
in this paper). The results for $\sigma$ are in the upper panel, the
results for $h_3$ in the middle panel and the results for $h_4$ in the
lower panel. The results for $\langle v\rangle$ are not shown because
the correction for the radial velocity of the template star would
introduce an additional uncertainty.

It is clear that the true errors (black bars) are larger than the
conventional errors and the realistic errors. For all three kinematic
parameters, it is remarkable that the conventional errors show little
variation if different stellar templates are used, whereas the
differences in values for the "realistic" errors are larger.  Moreover,
in cases with large true errors, the "realistic" errors are closer to
these values than the conventional errors. It seems that the errors obtained
with this method can be used as indicators for template mismatch in
the sense that larger errors indicate a poorer fitting template,
whereas the conventional errors clearly cannot be used as such.

\section{Conclusions}\label{conc}

As the technical limitations on the quality of observed data are
diminishing, the numerical signal processing starts putting
limitations on the information content of the observations.
Therefore, it is important to have realistic estimates of the errors
on the data.

In this paper, we propose a new method to calculate error estimates on
kinematic parameters derived from spectroscopic observations. The
starting point is the realization that the data points in the 'reduced
and calibrated' spectral images do not follow a Poisson distribution
and are not independent from pixel to pixel. This implies that the
error estimates derived with classical tools (Monte Carlo simulations
with Gaussian noise or statistical interpretation of least squares
techniques) will underestimate the true errors in many cases.

The proposed method first determines the characteristics of the real
noise on the data.  For this, a flexible fitting method offering
enough degrees of freedom is used. The residual of the fit is
considered as observed noise.  It is shown that the power spectrum of
this observed noise can be very different from white noise. Realistic
error estimates are obtained through Monte Carlo simulations with
synthetic galaxy spectra, where the noise distribution follows the
same power spectrum as the observed noise. In this way, errors are
estimated taking the real characteristics of the noise into account,
instead of relying on a Gaussian (or Poisson) noise distribution.
Moreover, the proposed scheme can be applied in combination with
several methods that are currently used to derive kinematic parameters
from spectroscopic data.

The method was tested on spectroscopic observations of NGC~3258. In
this particular case, the realistic errors are almost a factor of 2
larger than the errors based on least squares statistics.  What is
most important, is that the realistic error estimates are more
consistent with the scatter among neighbouring data points in the
kinematic profiles.

Simulations with spectra containing Poisson noise and impulse noise
confirm that the proposed method offers error estimates that are
closer to the true errors than conventional error estimates.  Moreover, from
simulations with synthetic spectra it becomes clear that an
oversampling of the spectra results in an underestimate of the errors
on kinematic parameters when simple least squares statistics are
used. This may be an important consideration when kinematic parameters
are calculated from logarithmically rebinned spectra.  Although this
method is not able to calculate realistic errors on kinematic
parameters obtained with an ill matching template star, the realistic
errors do trace the situation of template mismatch, in the sense that
they become larger with increasing template mismatch, unlike the usual
error estimates.

\section*{Acknowledgments}

\appendix
\section[]{Cubic spline fit}

\begin{figure}\hbox{
\includegraphics[scale=.2]{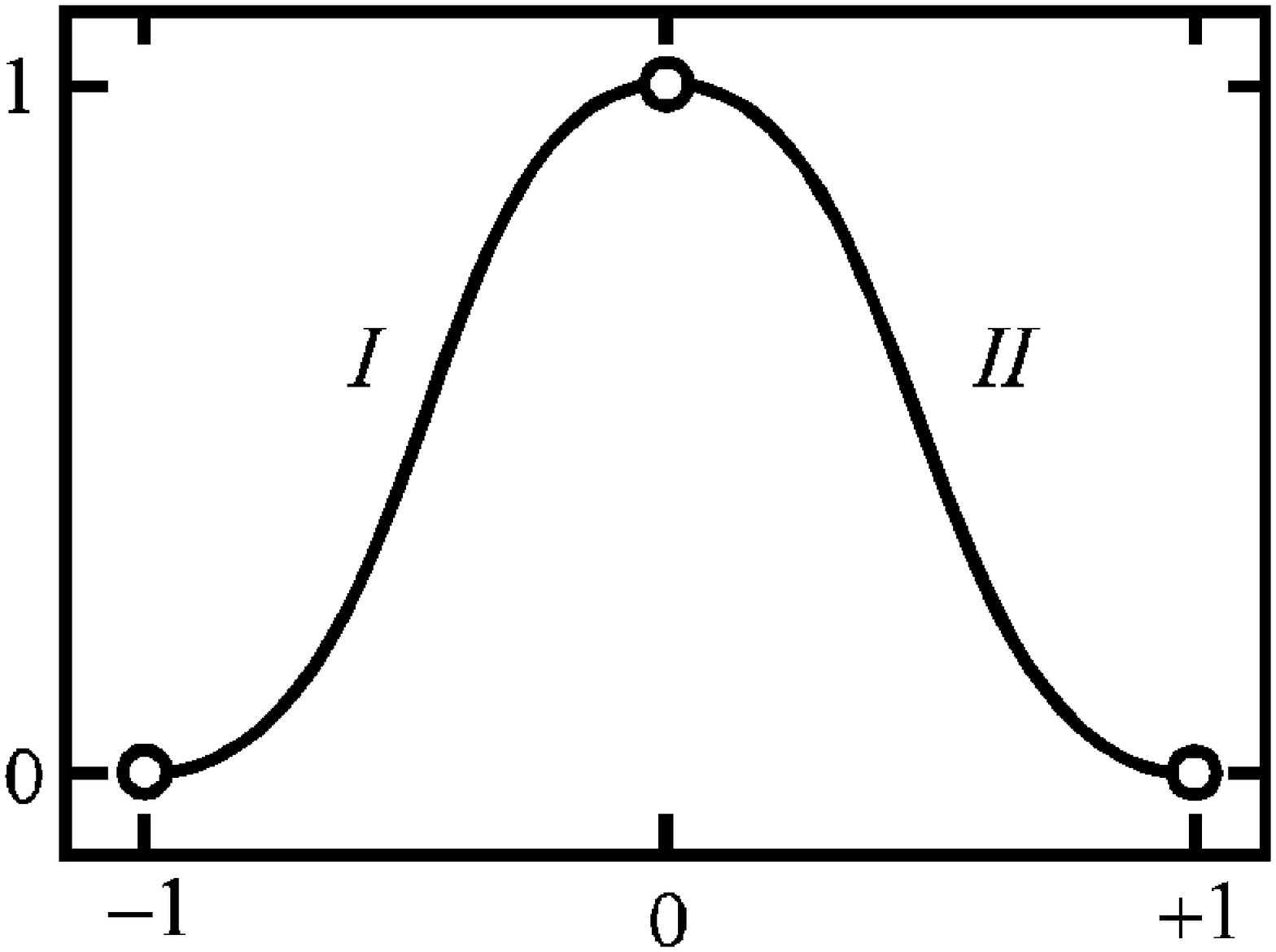}
\includegraphics[scale=.2]{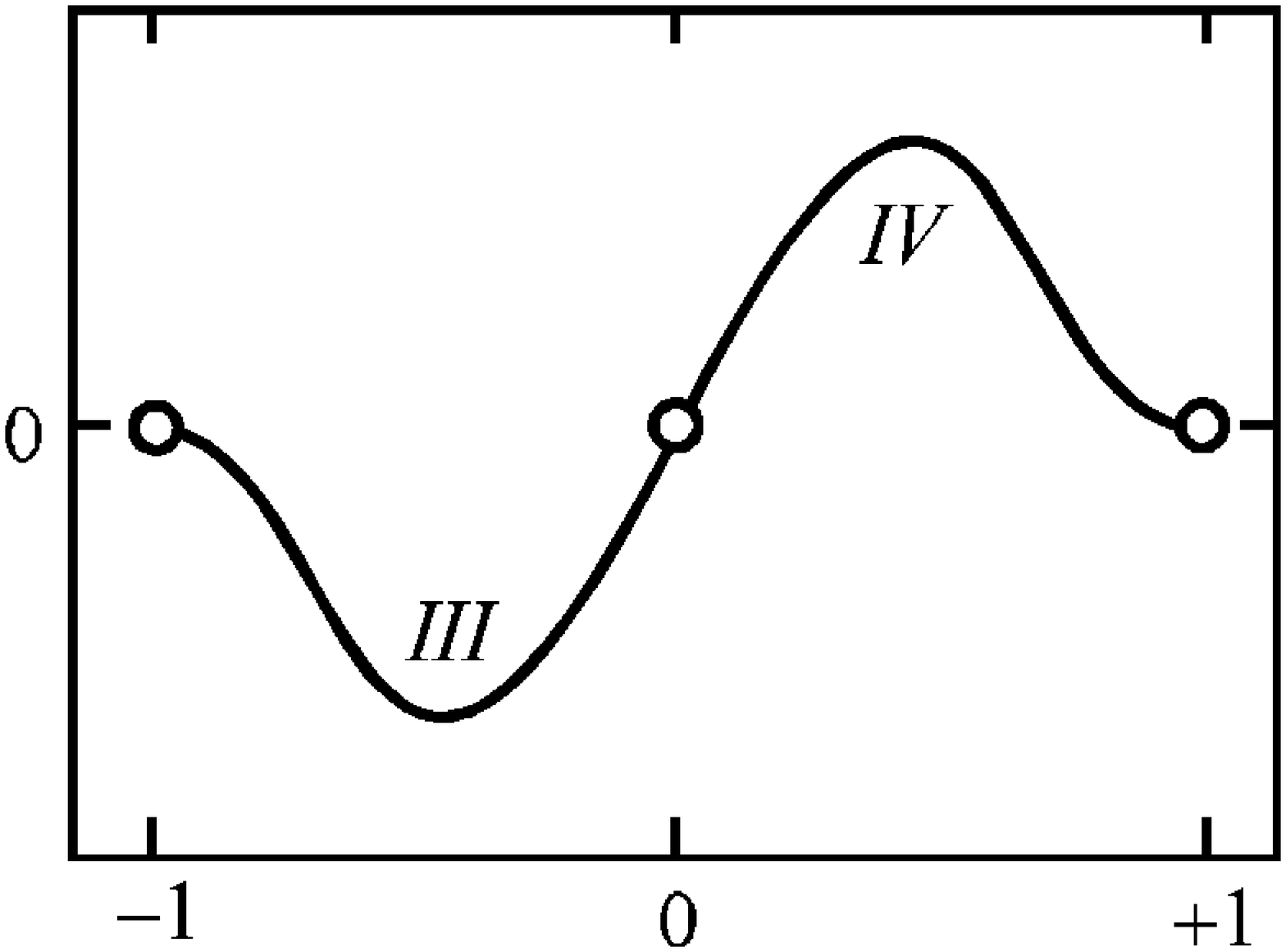}}

\caption{ Left: A first type of third degree cubic spline, composed
with the functions I and II. Right: A second type of third degree
cubic spline, composed with the functions III and IV. The definitions
can be found in the text.}

\label{qubics}
\end{figure}
In this paper, a new method is used to perform a non-parametric fit of
a model spectrum to an observed galaxy spectrum. The method performs a
least squares fit, using a set of cubic spline basis functions to
represent the LOSVD.

Each basis function is a third-degree cubic spline. There are two
types of cubic spline basis functions, each a composition of two
polynomials, as illustrated in figure \ref{qubics}.

In this figure, the cubic splines are defined in the interval
[-1,1]. The first third degree basis function (left in figure
\ref{qubics}) is a composition of polynomials I and II.  The second
basis function (right in figure \ref{qubics}) is a composition of
polynomials III and IV. These polynomials are:
\begin{eqnarray}
I&=&1-3x^2-2x^3=(1+x)^2(1-2x)\\
II&=& 1-3x^2+2x^3=(1-x)^2(1+2x)\\
III&=& x+2x^2+x^3=(1+x)^2x\\
IV&=& x-2x^2+x^3=-(1-x)^2x
\end{eqnarray}

Function I is defined in the interval [0,1] and meets the requirements
that $y(1)=0=y'(1)$, $y(0)=1$, $y'(0)=0$, (prime denoting first order
derivative). This behaviour is mirrored with respect to the vertical
axis into the region [-1,0], resulting in function II. The composition
of I and II gives a symmetric cubic spline basis function.  The first
two conditions assure that the function declines smoothly to 0 at the
edges. The other two conditions give the profile a flat peak on the
y-axis.

Function IV is defined in the interval [0,1] and meets the
requirements that $y(1)=0=y'(1)$, $y(0)=0$, $y'(0)=1$. This behaviour is
mirrored with respect to the centre into the region [-1,0], resulting
in function III. The composition of III and IV is an antisymmetric
cubic spline basis function. Again, the first two conditions assure
that the function declines smoothly to 0 at the edges. The other
conditions make sure the function goes through the centre and give a
steeply declining or inclining profile near the centre.

\begin{figure}
\includegraphics[scale=.4]{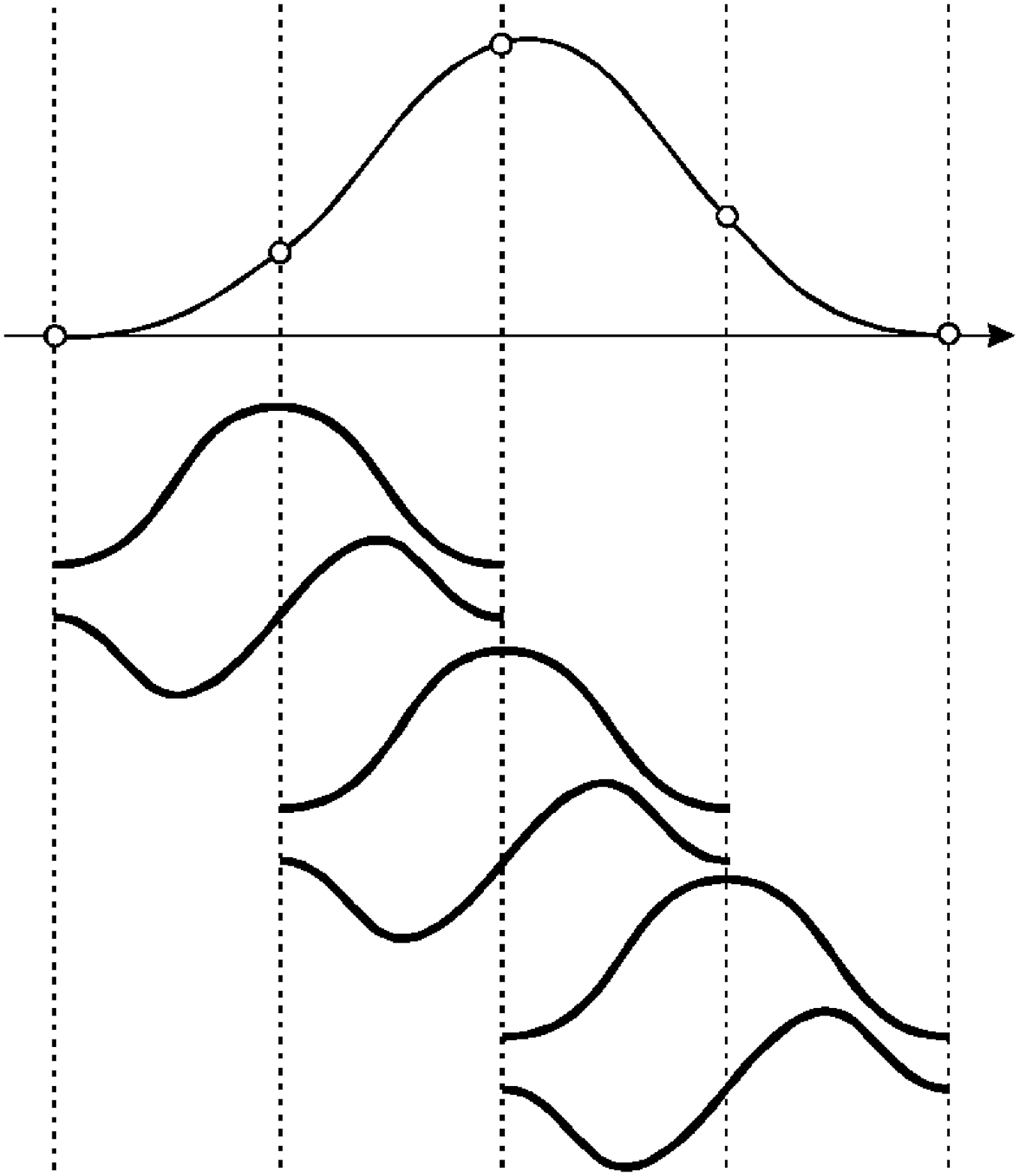}

\caption{ Implementation scheme for the non-parametric fit. The
fitting interval is divided in a number of sub-intervals. Every two
adjacent sub-intervals are used to define a set of cubic splines that
is included in the fit. This means that four basis functions of
different type (I, II, III, IV) are defined in each sub-interval.}

\label{impl}
\end{figure}

For the practical implementation, the interval where the fit is
performed is divided in $(n-1)$ sub-intervals, using $p_i$,
$i=1,\ldots,n$ points.  This is illustrated in figure \ref{impl}. Each
triplet of adjacent points $p_{i-1},p_i,p_{i+1}$ is used to define a set of
cubic splines that is included in the fit.  

\begin{figure}
\includegraphics[scale=.5]{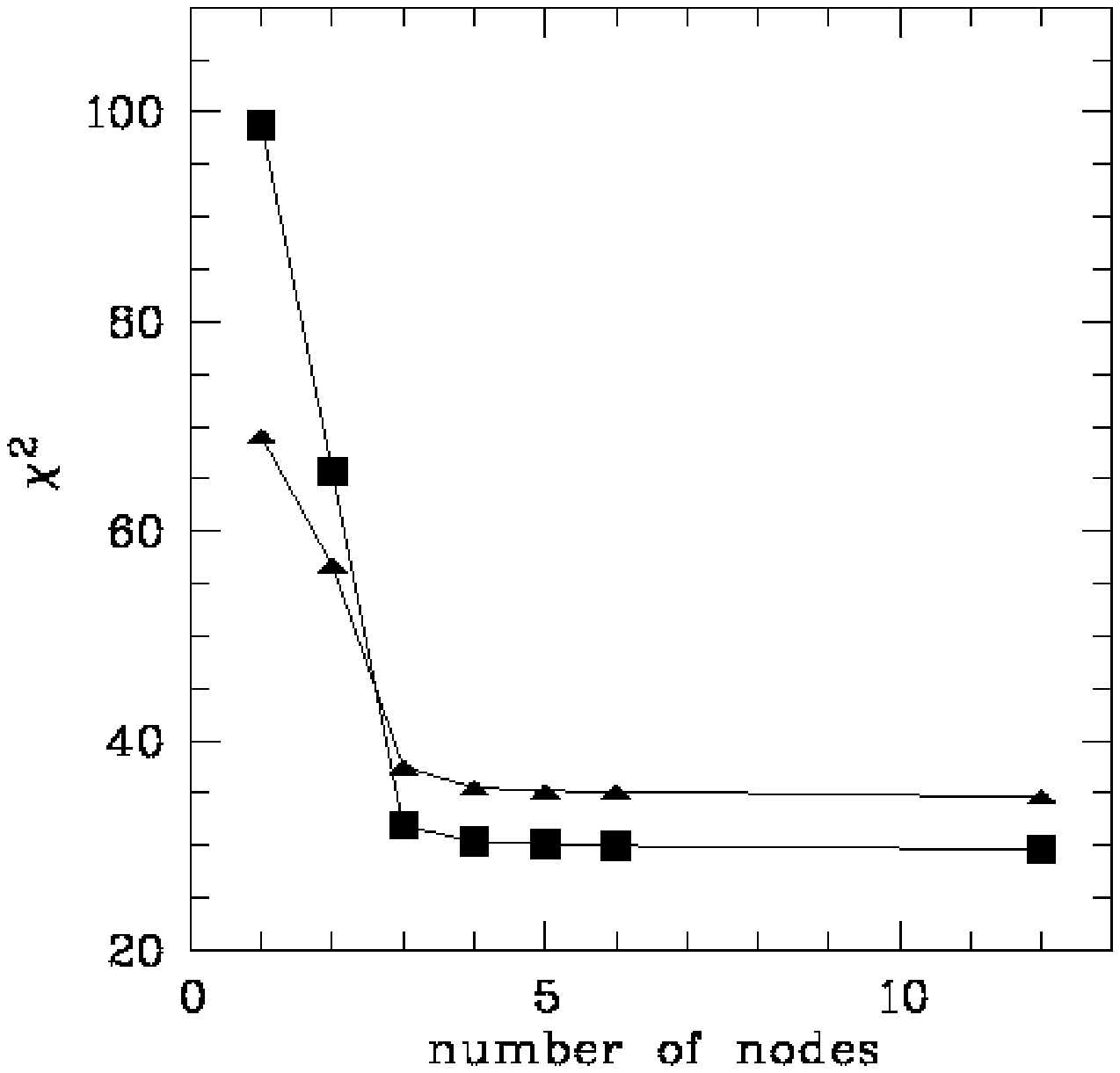}
\caption{The $\chi^2$ values of a number of fits with an increasing
number of sub-intervals, for the spectra at $r=0.8''$ (filled squares)
and $r=11''$ (filled triangles). For both spectra, using four nodes or
more gives about the same values for $\chi^2$.}
\label{chikes}
\end{figure}

The choice of the number of sub-intervals or basis functions is
free. The less sub-intervals are considered, the more the line profile
is smoothed. There is no a priori criterion that can be used to decide
how many of these sub-intervals have to be used to obtain a good
fit. Instead, a number of fits with an increasing number of
sub-intervals (hence basis functions) is performed and the $\chi^2$
values of the fits then indicate when the number of cubic spline pairs
offers enough degrees of freedom to come to a good fit. This is
illustrated in figure \ref{chikes}; once a sufficient degree of
freedom is reached, the values for $\chi^2$ start decreasing only very
slowly when the number of nodes is increased.  If $n$ points are used,
$(n-1)$ sub-intervals are used for the fit, yielding $2n-4$ degrees of
freedom.

Kuijken \& Merrifield (1993) came up with a similar idea, but they
used Gaussian functions for the decomposition of the LOSVD.  The cubic
splines used here are for the same degrees of freedom slightly more
flexible in fitting.
The fit is not restricted to strictly positive results. If one wants
to obtain physical solutions with this method, linear constraints can
be added to the $\chi^2$ fit.

\bsp

\label{lastpage}

\end{document}